\documentclass[journal]{IEEEtran}
\usepackage{amssymb,amsmath}
\usepackage{stmaryrd}
\usepackage{tikz}
  \usetikzlibrary{automata,shapes,backgrounds}
  \usetikzlibrary{decorations.pathmorphing}
\usepackage[linesnumbered,ruled,vlined]{algorithm2e}
\usepackage{array}
\usepackage{cite}
\usepackage{url}
\usepackage{enumitem}
\newcommand{\eps}{\varepsilon}
\newtheorem{thm}{Theorem}

\newtheorem{remark}[thm]{Remark}

\newtheorem{example}[thm]{Example}
\DeclareMathOperator{\A}{\mathcal A}
\DeclareMathOperator{\B}{\mathcal B}
\DeclareMathOperator{\D}{\mathcal D}

\begin{document}
\title{Complexity of Verifying Nonblockingness in Modular Supervisory Control\thanks{Supported by the DFG in Emmy Noether grant KR~4381/1-1 (DIAMOND).}}

\author{Tom{\' a}{\v s}~Masopust\thanks{T. Masopust ({\tt tomas.masopust{@}tu-dresden.de}) is with the Institute of Theoretical Computer Science and cfaed, TU Dresden, Germany}}

\markboth{}{}

\maketitle

\begin{abstract}
  Complexity analysis becomes a common task in supervisory control. However, many results of interest are spread across different topics. The aim of this paper is to bring several interesting results from complexity theory and to illustrate their relevance to supervisory control by proving new nontrivial results concerning nonblockingness in modular supervisory control of discrete event systems modeled by finite automata.
\end{abstract}


\section{Introduction}\label{intro}
  Nonblockingness 
  is an important property of discrete event systems ensuring that every task can be completed. It has therefore been intensively studied in the literature~\cite{MohajeraniMF14}. An automaton (deterministic or nondeterministic) is {\em nonblocking\/} if every sequence of events generated by the automaton can be extended to a marked sequence. Given a set of nonblocking automata, the {\em modular nonblockingness problem\/} asks whether the parallel composition of all the automata of the set results in a nonblocking automaton.
  
  The property is easy to verify for a deterministic automaton (DFA) as we discuss in Theorem~\ref{thm1}. However, if the automaton is nondeterministic (NFA) or a set of nonblocking DFAs is considered, the verification becomes computationally more demanding. We study the complexity in Theorems~\ref{pspace-complete02} and~\ref{pspace-complete01}, respectively. A result relevant to timed discrete event systems is provided in Theorem~\ref{np-unary}.

  So far, no efficient (polynomial) algorithm for verifying modular nonblockingness is known. In the light of the results of this paper, it is not surprising. The problem is complete for the complexity class for which the experts believe that no efficient algorithms exist. Therefore it is unlikely that there is an efficient algorithm solving the problem in general. However, there can still be optimization methods or algorithms working well for most of the practical cases. For instance, Malik~\cite{Malikwodes2016} has recently shown that explicit model checking algorithms without any special data structures work well on standard computers for several practical systems with 100 million states.

  The aim of this paper is to bring and apply some of the interesting results from automata and complexity theory to the nonblockingness verification problem.

\section{Preliminaries}\label{preliminaries}
  An {\em alphabet}, $\Sigma$, is a finite nonempty set. The elements of an alphabet are called {\em events}. A {\em string} over $\Sigma$ is a finite sequence ({\em catenation}) of events, for example, $001$ is a string over $\{0,1\}$. Let $\Sigma^*$ denote the set of all finite strings over $\Sigma$; the {\em empty string\/} is denoted by $\varepsilon$. 
  
  A {\em nondeterministic finite automaton\/} (NFA) over an alphabet $\Sigma$ is a structure $\A = (Q,\Sigma,\delta,I,F)$, where $Q$ is the finite nonempty set of states, $I\subseteq Q$ is the nonempty set of initial states, $F \subseteq Q$ is the set of accepting (marked) states, and $\delta \colon Q\times\Sigma \to 2^Q$ is the transition function that can be extended to the domain $2^Q\times\Sigma^*$ by induction. The {\em language generated by $\A$\/} is the set $L(\A) = \{w\in \Sigma^* \mid \delta(I,w)\neq\emptyset\}$ and the {\em language marked by $\A$\/} is the set $L_m(\A) = \{w\in \Sigma^* \mid \delta(I,w)\cap F \neq\emptyset\}$. Equivalently, the transition function is a relation $\delta \subseteq Q\times \Sigma \times Q$. Then the meaning of $\delta(q,a)=\{s,t\}$ is that there are two transitions $(q,a,s)$ and $(q,a,t)$.

  The prefix closure of a language $L$ is the set $\overline{L}=\{w\in \Sigma^* \mid \text{there exists } u \in\Sigma^* \text{ s.t. } wu\in L\}$; $L$ is {\em prefix-closed\/} if $L=\overline{L}$. Obviously, $L_m(\A)\subseteq L(\A)$ and $L(\A)$ is prefix-closed.

  NFA $\A$ is {\em deterministic\/} (DFA) if it has a unique initial state, $|I|=1$, and no nondeterministic transitions, $|\delta(q,a)|\le 1$ for every $q\in Q$ and $a \in \Sigma$. For DFAs, we identify singletons with their elements and simply write $p$ instead of $\{p\}$. Specifically, we write $\delta(q,a)=p$ instead of $\delta(q,a)=\{p\}$. 

  For every NFA $\A$ there exists a DFA $\B$ such that $L_m(\B)=L_m(\A)$ and $L(\B)=L(\A)$.
  
  Let $\Sigma$ and $\Gamma$ be alphabets, and let $f\colon\Sigma^* \to \Gamma^*$ be a map. Then $f$ is a {\em morphism} (for catenation) if $f(xy)=f(x)f(y)$ for every $x,y\in\Sigma^*$. Let $\Sigma_o\subseteq \Sigma$ be alphabets. A {\em projection} $P$ from $\Sigma^*$ to $\Sigma_o^*$ is a morphism defined by $P(a) = \varepsilon$ for $a\in \Sigma\setminus \Sigma_o$, and $P(a)= a$ for $a\in \Sigma_o$.
  The action of projection $P$ on a string $w \in \Sigma^*$ is to erase all events from $w$ that do not belong to $\Sigma_o$. The {\em inverse image} of $P$, denoted $P^{-1}$, is defined as $P^{-1}(s)=\{w\in \Sigma^* \mid P(w) = s\}$. The definitions can readily be extended to languages. 

  Let $L_i$ be a language over $\Sigma_i$, $i=1,\ldots,n$. The {\em parallel composition\/} of $(L_i)_{i=1}^{n}$ is defined by $\|_{i=1}^{n} L_i = \cap_{i=1}^{n} P_i^{-1}(L_i)$, where $P_i$ is a projection from $(\cup_{i=1}^{n} \Sigma_i)^*$ to $\Sigma_i^*$. 
  For $i=1,2$, let $\A_i=(Q_i,\Sigma_i,\delta_i,I_i,F_i)$ be NFAs. The {\em parallel composition of $\A_1$ and $\A_2$\/} is defined as the accessible part of the NFA $(Q_1\times Q_2, \Sigma_1\cup\Sigma_2, \delta, I_1\times I_2, F_1\times F_2)$, where 
  \[ \delta((x,y),e) = 
    \left\{ 
      \begin{array}{r@{\,}l@{\,}ll}
        \delta_1(x,e) & \times & \delta_2(y,e) & \text{if } e\in \Sigma_1\cap\Sigma_2\\
        \delta_1(x,e) & \times & \{y\} & \text{if } e\in \Sigma_1\setminus\Sigma_2\\
        \{x\}         & \times & \delta_2(y,e) & \text{if } e\in \Sigma_2\setminus\Sigma_1
      \end{array}
    \right.
  \]
  The parallel composition of DFAs is a DFA~\cite{CL08}.
  The relationship between the definitions is $L(\A_1 \| \A_2) = L(\A_1) \parallel L(\A_2)$ and $L_m(\A_1 \| \A_2) = L_m(\A_1) \parallel L_m(\A_2)$.

  An NFA $\A$ is {\em nonblocking\/} if $\overline{L_m(\A)} = L(\A)$. The inclusion $\overline{L_m(\A)} \subseteq L(\A)$ always holds.
  
\begin{figure}
      \centering
      \begin{tikzpicture}[baseline,auto,->,>=stealth,shorten >=1pt,node distance=1cm,
        state/.style={ellipse,minimum size=0mm,very thin,draw=black,initial text=},
        every node/.style={font=\small}]
        \node[state,initial,accepting]  (1) {};
        \node[state,accepting]          (2) [right of=1]  {};
        \path
          (1) edge node {$a$} (2)
          ;
      \end{tikzpicture}
      \quad
      \begin{tikzpicture}[baseline,auto,->,>=stealth,shorten >=1pt,node distance=1cm,
        state/.style={ellipse,minimum size=0mm,very thin,draw=black,initial text=},
        every node/.style={font=\small}]
        \node[state,initial,accepting]  (1) {};
        \node[state]                    (2) [right of=1]  {};
        \node[state,accepting]          (3) [right of=2]  {};
        \path
          (1) edge node {$a$} (2)
          (2) edge node {$a$} (3)
          ;
      \end{tikzpicture}
      \quad
      \begin{tikzpicture}[baseline,auto,->,>=stealth,shorten >=1pt,node distance=1cm,
        state/.style={ellipse,minimum size=0mm,very thin,draw=black,initial text=},
        every node/.style={font=\small}]
        \node[state,initial,accepting]  (1) {};
        \node[state]                    (2) [right of=1]  {};
        \path
          (1) edge node {$a$} (2)
          ;
      \end{tikzpicture}
      \caption{DFAs $\A_1$ and $\A_2$ and their parallel composition $\A_1 \| \A_2$}
      \label{fig2}
    \end{figure}
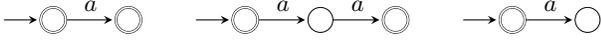

  To show that a composition of nonblocking automata can be blocking, let $\A_1$ and $\A_2$ be DFAs over $\{a\}$ depicted in Fig.~\ref{fig2}. Both automata are nonblocking but their parallel composition is blocking, because $a$ cannot be extended to a marked string.
  
  We now briefly recall the basic notions of complexity theory. For all unexplained notions, the reader is referred to the lit\-er\-a\-ture~\cite{papadimitriou,sipser}. 

  There are two complexity measures: space and time. The class NSPACE($f(n)$) denotes the class of all problems decidable by a nondeterministic Turing Machine (TM) (a nondeterministic algorithm) in space $O(f(n))$ for an input of size $n$. The class $\text{NL}=\text{NSPACE}(\log n)$ is thus the class of all problems decidable by a nondeterministic TM in logarithmic space, and $\text{PSPACE}=\text{NPSPACE}=\cup_{k\in\mathbb{N}}\text{NSPACE}(n^k)$ is the class of all problems decidable by a (nondeterministic) TM in polynomial space. The space required to store the input and output is not considered in space complexity.
  
  The class P (NP) denotes the class of all problems decidable by a (nondeterministic) TM in polynomial time. 
  
  The hierarchy of classes is $\textrm{NL} \subseteq \textrm{P} \subseteq \textrm{NP} \subseteq \textrm{PSPACE}$. Even though $\textrm{NL}\subsetneq \textrm{PSPACE}$, the strictness of any other inclusion is unknown. The (non)strictness of these inclusions is the most interesting and important open problem of complexity theory. 
  
  The classes NL and NP are defined in terms of a nondeterministic TM (a nondeterministic algorithm). Although for every nondeterministic TM there is an equivalent deterministic TM, the difference is in complexity. A typical nondeterministic step of a nondeterministic algorithm is ``choose $x \in X$''. Deterministically, one can imagine to check all the possibilities for $x$ one by one. Nondeterministically, the situation is different. There are two basic views how a nondeterministic algorithm performs a nondeterministic step.
  The first view is that the algorithm ``guesses'' the right value of $x$ that eventually leads to a success, if such a value exists. The other view is that the algorithm makes a copy of itself for every nondeterministic step with different value of $x$ in every copy. For a nondeterministic step ``choose $x\in\{1,\ldots,100\}$'', 100 copies of the algorithm would be created, where the value of $x$ in the $i$th copy is $x=i$. The nondeterministic algorithm is successful if at least one of the copies is successful. In this view, the time (space) complexity is the maximum of time (space) required by a copy.
  
  \begin{example}\label{ex3}
    Let $G=(V,E,s,t)$ be a directed graph with $s,t\in V$ the source and target nodes. The {\em graph reachability\/} problem asks whether the target node $t$ is reachable from the source node $s$ in $G$. The problem belongs to NL~\cite{papadimitriou}. 
    
    To show this, we describe a nondeterministic algorithm (Algorithm~\ref{alg1}) that solves the graph reachability problem in logarithmic space. Algorithm~\ref{alg1} is nondeterministic because of the nondeterministic step on line~\ref{nondetStep}. Following the first view, the algorithm correctly guesses the edges that lead from node $s$ to node $t$, if such a path exists. Following the second view, the algorithm forks for every possible edge on line~\ref{nondetStep}. If any of the copies ever reaches $t$, the copy returns {\tt true}, which is then the overall answer. The variable $numSteps$ counts the number of steps and terminates the cycle if it is bigger than the number of nodes. This is fine because if there is a path from $s$ to $t$, then there is a path of length at most $|V|-1$.

    \begin{algorithm}
      \DontPrintSemicolon
      \caption{(Graph reachability)}
      \label{alg1}
      \SetKwInOut{Input}{Input}\SetKwInOut{Output}{Output}
        \Input{A directed graph $G=(V,E,s,t)$} 
        \Output{{\tt true} iff $t$ is reachable from $s$ in $G$}
        $k := s$;
        $numSteps := 0$\;
        \Repeat{$k = t$ {\bf or} $numSteps > |V|-1$}{ 
          { choose $k'$ such that} $(k,k') \in E$\;\label{nondetStep}
          $k := k'$;
          $numSteps := numSteps + 1$\;
        }
        \lIf {$k=t$ } { \Return {\tt true} }
        \Return {\tt false}
    \end{algorithm}

    It remains to show that Algorithm~\ref{alg1} works in logarithmic space. Since the input is not considered in space complexity, the only space required by the algorithm is the space to store $k,k',|V|-1$ and $numSteps$. However, $numSteps$ is a binary number, bounded by $|V|$, which requires at most $\lceil\log |V|\rceil$ digits. Similarly, $k$ is a pointer to the position in the input, where the actual value of $k$ is stored. Thus, it is again a binary number with at most $\lceil\log(|V|+|E|)\rceil$ digits. Similarly for $k'$ and $|V|-1$.
  \hfill\IEEEQEDopen\end{example}

  A problem is PSPACE-complete if it can be solved using only polynomial space ({\em membership\/} in PSPACE) and if every problem that can be solved in polynomial space can be reduced (transformed) to it in polynomial time (PSPACE-{\em hardness}). PSPACE-complete problems are therefore the hardest problems in PSPACE. Similarly for the other complexity classes, with only a different requirement on the reduction. Namely, to prove NL-hardness, the reduction has to be in deterministic logarithmic space, and to prove NP-hardness, the reduction has to be in polynomial time (as well as for PSPACE-hardness).
  
  For instance, satisfiability of formulae in conjunctive normal form\footnote{\label{ft1}A (boolean) formula consists of variables, operators conjunction, disjunction and negation, and parentheses. A formula is satisfiable if there is an assigning of $1$ ({\it true}) and $0$ ({\it false}) to its variables making it {\it true}. A literal is a variable or its negation. A clause is a disjunction of literals. A formula is in 3-cnf if it is a conjunction of clauses, each clause with three literals. For instance, $\varphi = (x\lor y \lor z) \land (\neg x\lor y \lor z)$ is a formula in 3-cnf with two clauses $x\lor y \lor z$ and $\neg x\lor y \lor z$. Given a formula in 3-cnf, the 3CNF problem asks whether the formula is satisfiable. The formula $\varphi$ is satisfiable for, e.g., $(x,y,z)=(0,1,0)$.}
  (3CNF) is an NP-complete problem~\cite{GareyJ79}. Therefore, by definition, any problem in NP can be reduced to 3CNF in polynomial time. We show in Theorem~\ref{np-unary} that the {\em One-shared-event DFA modular nonblockingness} (1SE-DFA-MN) problem is in NP, hence reducible to 3CNF in polynomial time. 
  
  The membership in NP gives an upper bound on the complexity of 1SE-DFA-MN, which can still be polynomially or even trivially solvable. 
  To rule out this possibility, we further show that 1SE-DFA-MN is NP-hard (and hence NP-complete) by reducing 3CNF to 1SE-DFA-MN. Then, consequently, any problem in NP can be reduced to the 1SE-DFA-MN problem in polynomial time. Hence, from the complexity point of view, both problems are equally difficult.

\section{Complexity of Nonblockingness}
  Let $\A=(Q,\Sigma,\delta,I,F)$ be an NFA. We define the size of $\A$ as $|\A| = |Q|+|\Sigma|+|\delta|+|I|+|F|$. 

  A DFA is nonblocking iff from every state a marked state is reachable (in other words, every state is reachable and co-reachable). This property can be tested in linear time using the computation of strongly connected components~\cite{Tarjan72}. 
  From the complexity point of view, under the assumption that $\textrm{NL}\neq\textrm{P}$, a stronger result can be shown.
  
  \begin{thm}[DFA-nonblockingness]\label{thm1}
    Given a DFA $\A$, the problem whether $\overline{L_m(\A)} = L(\A)$ is NL-complete.
  \end{thm}
  \begin{IEEEproof}
    The membership of DFA-nonblockingness in NL follows from Algorithm~\ref{alg3} below for $n=1$.
    
    We now show that DFA-nonblockingness is NL-hard by reducing {\em graph non-reachability}~\cite{papadimitriou} to DFA-non\-blockingness. Namely, let $G=(V,E,s,t)$ be a directed graph with $s,t$ in $V$. We construct a DFA $\A$ from $G$ in logarithmic space such that $t$ is not reachable from $s$ in $G$ iff $\A$ is nonblocking.
    
    Let $\A=(V\cup\{t'\},\Sigma,\delta,s,V)$, where $\delta$ is defined as the relation $E$ with every transition under a unique label, and a transition under a new label is added from $t$ to the new non-marked state $t'$. This reduction (transformation) of $G$ to $\A$ can be done in logarithmic space and is performed by Algorithm~\ref{alg2}, where $\Sigma=\{1,2,\ldots,|E|+1\}$.
    If the algorithm reads a node $v$ in $V$, it outputs state $v$. Then it prints state $t'$. After this part, it has printed the state set of $\A$. It only needs to store a pointer (of logarithmic size) to the position of the input currently read. Then the algorithm counts from $1$ to $|E|+1$ and outputs the numbers, that is, it prints the alphabet of $\A$. For this, two numbers, $i$ and $|E|+1$ with at most $\lceil \log(|E|+1) \rceil$ digits are stored. Then, it reads the input again (using the pointer as above) and uses a counter $c$ (with at most $\lceil \log(|E|+1) \rceil$ digits) to print, for every edge $(u,v)$ in $E$, the corresponding transition $(u,c,v)$ of $\delta$. Finally, it prints the transition $(t,|E|+1,t')$, state $s$, and all $v\in V$. After this, the output contains the DFA $\A$. The reduction uses logarithmic space to produce the output. Recall that the size of the input and output is not considered in space complexity.

    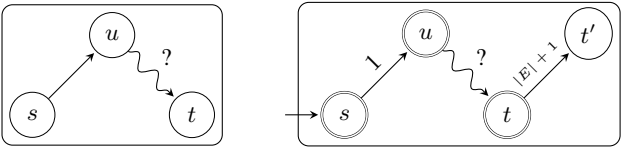
\begin{figure}[t]
      \centering
      \begin{tikzpicture}[baseline,auto,->,>=stealth,shorten >=1pt,node distance=1.5cm,
        state/.style={ellipse,minimum size=6mm,very thin,draw=black,initial text=},
        every node/.style={font=\small}]
        \node[state]                    (1) {$s$};
        \node[state]                    (2) [above right of=1]  {$u$};
        \node[state]                    (3) [below right of=2]  {$t$};
        \path
          (1) edge node {} (2)
          (2) edge[style={decorate, decoration={snake,post length=1.3mm}}] node{?} (3) ;
          ;
        \begin{pgfonlayer}{background}
          \path (2.north -| 3.east) + (0.1,0.1)    node (a) {};
          \path (1.south -| 1.west) + (-0.1,-0.1)  node (b) {};
          \path[rounded corners, draw=black] (a) rectangle (b);
        \end{pgfonlayer}
      \end{tikzpicture}
      \quad
      \begin{tikzpicture}[baseline,auto,->,>=stealth,shorten >=1pt,node distance=1.53cm,
        state/.style={ellipse,minimum size=6mm,very thin,draw=black,initial text=},
        every node/.style={font=\small}]
        \node[state,initial,accepting]  (1) {$s$};
        \node[state,accepting]          (2) [above right of=1]  {$u$};
        \node[state,accepting]          (3) [below right of=2]  {$t$};
        \node[state]                    (4) [above right of=3]  {$t'$};
        \path
          (1) edge node[sloped,above] {$1$} (2)
          (3) edge node[pos=0.5,sloped,above] {{\tiny $|E|+1$}} (4)
          (2) edge[style={decorate, decoration={snake,post length=1.3mm}}] node{?} (3) ;
          ;
        \begin{pgfonlayer}{background}
          \path (2.north -| 4.east) + (0.1,0.1)    node (a) {};
          \path (1.south -| 1.west) + (-0.3,-0.1)  node (b) {};
          \path[rounded corners, draw=black] (a) rectangle (b);
        \end{pgfonlayer}
      \end{tikzpicture}
      \caption{Graph $G=(V,E,s,t)$ and its corresponding DFA $\A$ of Theorem~\ref{thm1}}
      \label{fig4}
    \end{figure}

    \begin{algorithm}
      \DontPrintSemicolon
      \caption{(Reduction of a graph to a DFA)}
      \label{alg2}
      \SetKwInOut{Input}{Input}\SetKwInOut{Output}{Output}
        \Input{A directed graph $G=(V,E,s,t)$}
        \Output{The DFA $\A=(V\cup\{t'\},\Sigma,\delta,s,V)$}
          \lFor {each $v\in V$} {output $v$}
          output $t'$\;
          \lFor {$i=1,\ldots, |E|+1$} {output  $i$}
          $c:=1$;
          \lFor {each $(u,v)\in E$ } {\{output $(u,c,v)$; $c\texttt{++}$\}}
          output $(t,c,t')$; output $s$\;
          \lFor {each $v\in V$} {output $v$}
    \end{algorithm}

    It is not difficult to see that $t$ is reachable from $s$ in $G$ iff $t$ is accessible from the initial state $s$ in $\A$. Namely, if $t$ is not accessible in $\A$, then all accessible states are marked and the language of $\A$ is nonblocking. If $t$ is accessible in $\A$, then so is $t'$, which is not marked and makes thus the language of $\A$ blocking; cf. Fig.~\ref{fig4} for an illustration.
  \end{IEEEproof}

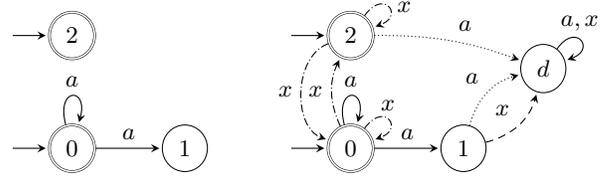
\begin{figure}
    \centering
    \begin{tikzpicture}[baseline,auto,->,>=stealth,shorten >=1pt,node distance=1.5cm,
      state/.style={ellipse,minimum size=6mm,very thin,draw=black,initial text=},
      every node/.style={font=\small}]
      \node[state,initial,accepting]  (1) {$0$};
      \node[state]                    (2) [right of=1] {$1$};
      \node[state,initial,accepting]  (3) [above of=1] {$2$};
      \path
        (1) edge node {$a$} (2)
        (1) edge[loop above] node {$a$} (1)
        ;
    \end{tikzpicture}
    \qquad
    \begin{tikzpicture}[baseline,auto,->,>=stealth,shorten >=1pt,node distance=1.5cm,
      state/.style={ellipse,minimum size=6mm,very thin,draw=black,initial text=},
      every node/.style={font=\small}]
      \node[state,initial,accepting]  (1) {$0$};
      \node[state]                    (2) [right of=1] {$1$};
      \node[state,initial,accepting]  (3) [above of=1] {$2$};
      \node[state]                    (d) [above right of=2] {$d$};
      \path
        (1) edge node {$a$} (2)
        (1) edge[loop above] node {$a$} (1)
        (2) edge[densely dotted,bend left]  node {$a$} (d)
        (2) edge[densely dashed,bend right] node {$x$} (d)
        (3) edge[densely dotted,bend left=10] node {$a$} (d)
        (d) edge[out=55,in=20,looseness=6] node[above] {$a,x$} (d)
        (3) edge[densely dash dot,out=55,in=20,looseness=6] node[right] {$x$} (3)
        (1) edge[densely dash dot,bend left=25] node[left] {$x$} (3)
        (3) edge[densely dash dot,bend right=70] node[left] {$x$} (1)
        (1) edge[densely dash dot,out=55,in=20,looseness=6] node[above] {$x$} (1)
        ;
    \end{tikzpicture}
    \caption{A nonblocking NFA $\B$ with $L_m(\B)=L(\B)=\{a\}^*$ and its cor\-re\-spond\-ing NFA $\A$ constructed in the proof of Theorem~\ref{pspace-complete02} }
    \label{fig3}
  \end{figure}

  Therefore, to check nonblockingness of a DFA is easy. This is, however, not true for NFAs. An NFA can be nonblocking even if there is a state from which no marked state is reachable, cf. Fig.~\ref{fig3}.

  \begin{thm}[NFA-nonblockingness]\label{pspace-complete02}
    Given an NFA $\A$, the problem whether $\overline{L_m(\A)} = L(\A)$ is PSPACE-complete.
  \end{thm}
  \begin{IEEEproof}
    To show that the problem is in PSPACE, let $\A = \allowbreak (Q,\Sigma,\delta,I,F)$ be an NFA. Let $\D$ be a DFA obtained from $\A$ by the standard subset construction~\cite{sipser}. States of $\D$ are subsets of states of $\A$, and $\A$ is nonblocking iff $\D$ is nonblocking. To check nonblockingness of $\D$ in polynomial space, $\D$ cannot be computed and stored, because it may require exponential space in the size of $\A$. Instead, we use the on-the-fly technique that keeps only a small part of $\D$ in memory and re-computes the required parts on request. Namely, for every state $X\subseteq Q$ of $\D$, we check that $X$ is reachable from the initial state of $\D$ (in the way depicted in Example~\ref{ex3}). If so, we guess a marked state $Y$ of $\D$, that is, $Y \cap F \neq\emptyset$, and check that $Y$ is reachable from $X$. This principle is generalized in Algorithm~\ref{alg3} below. At any time during the computation, the algorithm stores only a constant number of states of $\D$, which are subsets of the state set $Q$ of $\A$. Therefore, the algorithm uses space polynomial in the size of $\A$ and the problem is thus in PSPACE.

    To show that NFA-nonblockingness is PSPACE-hard, we reduce the NFA {\em universality problem}~\cite{GareyJ79} to it. NFA universality asks, given an NFA $\B$ over $\Sigma$, whether $L_m(\B) = \Sigma^*$.

    Let $\B=(Q,\Sigma,\delta_{\B},I,F)$ be an NFA, and let $d$ be a new non-marked state. We ``complete'' $\B$ in the sense that if, for an event $a$ in $\Sigma$, no $a$-transition is defined in a state $q$, we add the transition $(q,a,d)$ to the transition relation, see the dotted transitions in Fig.~\ref{fig3}. Let $x\notin\Sigma$ be a new event. State $d$ contains self-loops under all events of $\Sigma\cup\{x\}$. For each non-marked state $p$, we add the transition $(p,x,d)$, see the dashed transitions in Fig.~\ref{fig3}, while for each marked state $p$, we add the transitions $(p,x,i)$ for every initial state $i$ in $I$, cf. the dot-dash transitions in Fig.~\ref{fig3}. Let $\A=(Q\cup\{d\},\Sigma\cup\{x\},\delta_{\A},I,F)$ denote the resulting NFA. Notice that $L_m(\B)\subseteq L_m(\A)$. We now show that $\B$ is universal iff $\A$ is nonblocking.

    If $\B$ is universal, that is, $L_m(\B)=\Sigma^*$, we show that $\A$ is nonblocking by showing that $L_m(\A)=L(\A)$. It is sufficient to show that $L(\A)\subseteq L_m(\A)$. Let $w \in L(\A)$. We proceed by induction on the number of occurrences of event $x$ in $w$. If $x$ does not occur in $w$, then $w\in\Sigma^*=L_m(\B)\subseteq L_m(\A)$. Thus, assume that $w=w_1xw_2$ with $w_1 \in \Sigma^*$ and $w_2 \in (\Sigma\cup\{x\})^*$. Since $w_1 \in \Sigma^* = L_m(\B)\subseteq L_m(\A)$, there is a path in $\A$ from an initial state $i$ in $I$ to a marked state $f$ in $F$ labeled by $w_1$. By construction, $I \subseteq \delta_{\A}(i,w_1x)$, since $\delta_{\A}(f,x)=I$. By the induction hypothesis, $\delta_{\A}(I,w_2)\cap F \neq \emptyset$, hence $w \in L_m(\A)$. Thus, $L_m(\A)=L(\A)$, which means that $\A$ is nonblocking.

    If $\B$ is not universal, that is, $L_m(\B) \neq \Sigma^*$, then there exists a $w$ in $\Sigma^*$ such that $\delta_{\A}(I,w) \cap F = \emptyset$, since for any $w$ over $\Sigma$, $\delta_{\A}(I,w)\cap F=\delta_{\B}(I,w)\cap F$. By the construction, $\delta_{\A}(I,wx)=\{d\}$, from which no marked state is reachable. Since $\A$ is complete, $wx$ belongs to $L(\A)$, therefore $\A$ is blocking.
  \end{IEEEproof}

  The situation with NFAs is even worse as shown now.
  \begin{thm}[NFA-prefix-closed]\label{cor1}
    Given an NFA $\A$, the problem whether $L_m(\A)$ is prefix-closed is PSPACE-complete.
  \end{thm}
  \begin{IEEEproof}
    Let $\A$ be an NFA. Then $L_m(\A)=\overline{L_m(\A)}$ iff the DFA $\D$ obtained from $\A$ by the standard subset construction has no reachable and co-reachable non-marked states. Since the class $\textrm{PSPACE}$ is closed under complement, we can check the opposite -- a nondeterministic algorithm guesses a subset of non-marked states of $\A$ and verifies, using the on-the-fly technique, that they form a reachable and co-reachable state in $\D$. The NFA-prefix-closed problem is thus in PSPACE.

    To show PSPACE-hardness, Hunt~III and Rosenkrantz~\cite{HuntR78} have shown that a property $R$ of languages over $\{0,1\}$ such that $R(\{0,1\}^*)$ is true and there exists a regular language that is not expressible as a quotient $x\backslash L =\{ w \mid xw \in L\}$, for some $L$ for which $R(L)$ is true, is as hard as to decide ``$=\{0,1\}^*$''. Since prefix-closedness is such a property (the class of prefix-closed languages is closed under quotient) and universality is PSPACE-hard for NFAs, the result implies that the NFA-prefix-closed problem is PSPACE-hard.
  \end{IEEEproof}

  These results justify why the attention is mostly focused on DFAs rather than NFAs. In the rest of the paper, we also focus on DFAs, unless stated otherwise.

\subsection{Modular Nonblockingness Problem}
  We now focus on the modular nonblockingness problem.
  The simplest case is that there is no interaction between the different subsystems. The following result is well known.
  \begin{thm}
    Let $J$ be a finite set, and let $\A_j$ be a nonblocking NFA over $\Sigma_j$, for $j\in J$. If the alphabets are pairwise disjoint, that is, $\Sigma_j\cap\Sigma_{j'}=\emptyset$ for $j\neq j'$, then the parallel composition $\overline{L_m(\|_{j\in J} \A_j)} = L(\|_{j\in J} \A_J)$ is nonblocking.
    \hfill\IEEEQEDclosed
  \end{thm}

  In many complex systems, it is however the case that there are events shared between the subsystems. In such a case, non\-blockingness is in general PSPACE-complete~\cite{RohloffL05}. A more fine-grained complexity can be distinguished based on the following criteria. Let $(\A_i)_{i=1}^{n}$ be DFAs:
  \begin{enumerate}
    \item The number of DFAs is not restricted.
    \item The number of DFAs is restricted by a function $g(m)$, that is, $n\le g(m)$, where $m$ is the length of the encoding of the DFAs $\A_1,\A_2,\ldots,\A_n$.
    \item The number of DFAs is restricted by a constant $k$, that is, $n\le k$.
  \end{enumerate}
  Case~2 is the most general one and deserves a discussion. Assume, for example, that our encoding of $\A_i$ requires $c > 1$ bits and that the encoding of $\A_1, \ldots, \A_n$ requires $m=n\cdot c$ bits. If $g(m)=m$, then $n \le g(nc)$ for every $n\ge 1$, which results in the non-restricted case~1. If $g(m)=k$, for a constant $k$, then $n \le g(nc)$ iff $n\le k$, which results in the restriction of case~3. See also Remark~\ref{rem7} below.

  We can now prove the following result.
  \begin{thm}[DFA $g(m)$-bounded modular nonblockingness]\label{pspace-complete01}
    Given nonblocking DFAs $(\A_i)_{i=1}^{n}$ with $\A_i$ over $\Sigma_i$, $2\le n\le g(m)$, where $m$ is the length of an encoding of the sequence of DFAs $\A_1,\A_2,\ldots,\A_n$. The problem whether $\overline{L_m(\|_{i=1}^{n} \A_i)} = L(\|_{i=1}^{n} \A_i)$ is NSPACE$(g(m)\log m)$-complete.
  \end{thm}
  
    \begin{algorithm}
    \DontPrintSemicolon
      \caption{Is $\A = \|_{i=1}^{n} \A_i$ nonblocking?}
      \label{alg3}
      \SetKwInOut{Input}{Input}\SetKwInOut{Output}{Output}
      \Input{Encoding of $\A_1,\ldots,\A_n$ of size $m$}
      \Output{{\tt yes} iff $\|_{i=1}^{n} \A_i$ is nonblocking}
        \For {each $(p_1,\ldots,p_n)\in \vartimes_{i=1}^{n} Q_i$}{
          \If {$(p_1,\ldots,p_n)$ is reachable from the init st. of $\A$}{
            choose $(s_1,\ldots,s_n)\in \vartimes_{i=1}^{n} F_i$\;
            $k_i:= p_i$, for $i=1,2,\ldots,n$\;
            \Repeat {$k_i = s_i$, for $i=1,2,\ldots,n$}{
              { choose} $a\in\Sigma$\;
              $k_i := \delta(k_i,a)$, for $i=1,2,\ldots,n$
            }
          }
        }
        \Return {\tt yes}
    \end{algorithm}

  \begin{IEEEproof}
    Let $(\A_i)_{i=1}^{n}$ over $(\Sigma_i)_{i=1}^{n}$ be nonblocking DFAs. Algorithm~\ref{alg3} solves the $g(m)$-bounded modular nonblockingness problem.
    It works as follows: for every reachable state $(p_1,\ldots,p_n)$ of $\A$ (lines~1-2), the algorithm nondeterministically chooses a marked state $(s_1,\ldots,s_n)$ of $\A$ (line~3) that is reachable from state $(p_1,\ldots,p_n)$ (lines~4-8). The algorithm returns {\tt yes\/} iff there is such a marked state for every reachable state, hence iff $\A$ is nonblocking. (Compared to Example~\ref{ex3}, we omitted the counter $numSteps$ for simplicity. It should be clear how the counter is introduced to make the algorithm always terminate.)

    During the computation, the algorithm stores only a constant number of $n$-tuples of pointers $(t_1,\ldots,t_n)$. The space used is therefore $O(n \log m)$. Since $n$ is bounded by $g(m)$, the space used by Algorithm~\ref{alg3} is $O(g(m) \log m)$, hence the problem is in $\text{NSPACE}(g(m)\log m)$.
    
    To prove hardness, we reduce the $\text{NSPACE}(g(m)\log m)$-complete {\em finite DFA intersection problem (DFA-int)}~\cite{HolzerK11} to our problem. DFA-int asks, given DFAs $(\B_i)_{i=1}^{n}$ with $2\le n\le g(m)$, where $m$ is the length of the encoding of the sequence of $\B_1,\ldots,\B_n$, whether $\cap_{i=1}^{n} L_m(\B_i) = \emptyset$. The DFAs $\B_1,\ldots,\B_n$ are over a common alphabet $\Sigma$.
    
    We now describe a deterministic logarithmic-space reduction from DFA-int to DFA $g(m)$-bounded modular nonblockingness. Notice that $n\ge 2$. Let $x\notin\Sigma$ be a new event. 
    
    We construct $\A_1$ from $\B_1$ by adding two new states $d_1$ and $d_1'$ and $x$-tran\-si\-tions from every marked state of $\B_1$ to $d_1$, and from $d_1$ to $d_1'$, see an illustration in Fig.~\ref{fig6}. All states of $\A_1$, but $d_1$, are marked, that is, $L_m(\A_1)=L(\B_1)\cup L_m(\B_1)xx$ and $L(\A_1)=L(\B_1)\cup L_m(\B_1)\{x,xx\}$.

    For every $i\ge 2$, we construct $\A_i$ from $\B_i$ by adding a new state $d_i$ and $x$-transitions from every marked state of $\B_i$ to $d_i$. All states of $\A_i$ are marked, hence $L_m(\A_i)=L(\A_i)=L(\B_i)\cup L_m(\B_i)x$.
    
    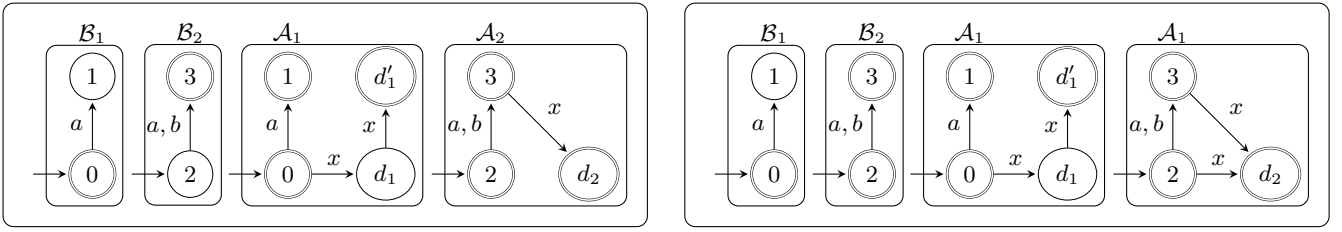
\begin{figure*}
      \centering
      \hfill
      \begin{tikzpicture}[baseline,auto,->,>=stealth,shorten >=1pt,node distance=1.3cm,
        state/.style={ellipse,minimum size=6mm,very thin,draw=black,initial text=},
        every node/.style={font=\small},framed, rounded corners]
        \node[state,initial,accepting]  (1) {$0$};
        \node[state]                    (2) [above of=1,label={[anchor=south]above:$\B_1$}] {$1$};
        \node[state,initial]            (3) [right of=1] {$2$};
        \node[state,accepting]          (4) [above of=3,label={[anchor=south]above:$\B_2$}] {$3$};
        \path
          (1) edge node {$a$} (2)
          (3) edge node {$a,b$} (4)
          ;
        \node[state,initial,accepting]  (11) [right of=3] {$0$};
        \node[state,accepting]          (12) [above of=11,label={[anchor=south]above:$\A_1$}] {$1$};
        \node[state]                    (d1) [right of=11] {$d_1$};
        \node[state,accepting]          (d') [above of=d1] {$d_1'$};
        \node[state,initial,accepting]  (13) [right of=d1,node distance=1.4cm] {$2$};
        \node[state,accepting]          (14) [above of=13,label={[anchor=south]above:$\A_2$}] {$3$};
        \node[state,accepting]          (d2) [right of=13] {$d_2$};
        \path
          (11) edge node {$a$} (12)
          (13) edge node {$a,b$} (14)
          (11) edge node {$x$} (d1)
          (d1) edge node {$x$} (d')
          (14) edge node {$x$} (d2)
          ;
        \begin{pgfonlayer}{background}
          \path (2.north -| 2.east) + (0.1,0.1)    node (a) {};
          \path (1.south -| 1.west) + (-0.3,-0.1)  node (b) {};
          \path[rounded corners, draw=black] (a) rectangle (b);
          \path (4.north -| 4.east) + (0.1,0.1)    node (a) {};
          \path (3.south -| 3.west) + (-0.3,-0.1)  node (b) {};
          \path[rounded corners, draw=black] (a) rectangle (b);
          \path (12.north -| d1.east) + (0.1,0.1)    node (a) {};
          \path (11.south -| 11.west) + (-0.3,-0.1)  node (b) {};
          \path[rounded corners, draw=black] (a) rectangle (b);
          \path (14.north -| d2.east) + (0.1,0.1)    node (a) {};
          \path (13.south -| 13.west) + (-0.3,-0.1)  node (b) {};
          \path[rounded corners, draw=black] (a) rectangle (b);
        \end{pgfonlayer}
      \end{tikzpicture}
      \hfill
      \begin{tikzpicture}[baseline,auto,->,>=stealth,shorten >=1pt,node distance=1.3cm,
        state/.style={ellipse,minimum size=6mm,very thin,draw=black,initial text=},
        every node/.style={font=\small}, framed, rounded corners]
        \node[state,initial,accepting]  (1) {$0$};
        \node[state]                    (2) [above of=1,label={[anchor=south]above:$\B_1$}] {$1$};
        \node[state,initial,accepting]  (3) [right of=1] {$2$};
        \node[state,accepting]          (4) [above of=3,label={[anchor=south]above:$\B_2$}] {$3$};
        \path
          (1) edge node {$a$} (2)
          (3) edge node {$a,b$} (4)
          ;
        \node[state,initial,accepting]  (11) [right of=3] {$0$};
        \node[state,accepting]          (12) [above of=11,label={[anchor=south]above:$\A_1$}] {$1$};
        \node[state]                    (d1) [right of=11] {$d_1$};
        \node[state,accepting]          (d') [above of=d1] {$d_1'$};
        \node[state,initial,accepting]  (13) [right of=d1,node distance=1.4cm] {$2$};
        \node[state,accepting]          (14) [above of=13,label={[anchor=south]above:$\A_1$}] {$3$};
        \node[state,accepting]          (d2) [right of=13] {$d_2$};
        \path
          (11) edge node {$a$} (12)
          (13) edge node {$a,b$} (14)
          (11) edge node {$x$} (d1)
          (d1) edge node {$x$} (d')
          (13) edge node {$x$} (d2)
          (14) edge node {$x$} (d2)
          ;
        \begin{pgfonlayer}{background}
          \path (2.north -| 2.east) + (0.1,0.1)    node (a) {};
          \path (1.south -| 1.west) + (-0.3,-0.1)  node (b) {};
          \path[rounded corners, draw=black] (a) rectangle (b);
          \path (4.north -| 4.east) + (0.1,0.1)    node (a) {};
          \path (3.south -| 3.west) + (-0.3,-0.1)  node (b) {};
          \path[rounded corners, draw=black] (a) rectangle (b);
          \path (12.north -| d1.east) + (0.1,0.1)    node (a) {};
          \path (11.south -| 11.west) + (-0.3,-0.1)  node (b) {};
          \path[rounded corners, draw=black] (a) rectangle (b);
          \path (14.north -| d2.east) + (0.1,0.1)    node (a) {};
          \path (13.south -| 13.west) + (-0.3,-0.1)  node (b) {};
          \path[rounded corners, draw=black] (a) rectangle (b);
        \end{pgfonlayer}
      \end{tikzpicture}
      \hfill
      \caption{DFAs $\B_1$ and $\B_2$ over $\Sigma=\{a,b\}$ and the corresponding DFAs $\A_1$ and $\A_2$ of Theorem~\ref{pspace-complete01}; a nonblocking instance (left) with $\overline{L_m(\A_1\|\A_2)}=L(\A_1 \| \A_2) = \{\eps,a\}$ and a blocking instance (right) with $\overline{L_m(\A_1\|\A_2)}=\{\eps,a\}$ and $L(\A_1 \| \A_2) = \{\eps,a,x\}$}
      \label{fig6}
    \end{figure*}
    
    We thus have that
    $
      L_m(\|_{i=1}^{n} \A_i) = \cap_{i=1}^{n} L(\B_i)
    $
    and
    \[
      L(\|_{i=1}^{n} \A_i) = \cap_{i=1}^{n}  L(\B_i) \cup \cap_{i=1}^{n}  L_m(\B_i)x\,. 
    \]
    We show $\overline{L_m(\|_{i=1}^{n} \A_i)} = L(\|_{i=1}^{n} \A_i)$ iff $\cap_{i=1}^{n} L_m(\B_i) = \emptyset$. 
    
    If $\cap_{i=1}^{n} L_m(\B_i) = \emptyset$, then $L(\|_{i=1}^{n} \A_i) = \cap_{i=1}^{n}  L(\B_i) = L_m(\|_{i=1}^{n} \A_i)$. 
    
    If $\cap_{i=1}^{n} L_m(\B_i) \neq \emptyset$, then there exists $w\in \cap_{i=1}^{n} L_m(\B_i)$, hence $wx\in L(\|_{i=1}^{n} \A_i) \setminus \overline{L_m(\|_{i=1}^{n} \A_i)}$, because $L_m(\|_{i=1}^{n} \A_i)$ does not contain any string with event $x$.
  \end{IEEEproof}

  \begin{remark}\label{rem7}
  Let $k$ be a constant. If for every $m$, $g(m)\le k$, then Algorithm~\ref{alg3} uses space $O(\log m)$, hence DFA $k$-bounded modular nonblockingness is in $\text{NL}=\text{NSPACE}(\log m)$ and it is NL-complete. If $g(m)\le \log^k m$, then DFA ($\log^k m$)-bounded modular nonblockingness is NSPACE$(\log^{k+1} m)$-complete. It is called a poly-logarithmic space complexity.
  \end{remark}

\subsection{One-Shared-Event Modular Nonblockingness}
  We now focus on the case where exactly one event is shared. An application of this case is, for example, in the Brandin and Wonham modular framework for timed discrete event systems~\cite{BrandinW1994}, where only one event simulating the tick of a global clock is shared and all the other events are local~\cite{SchafaschekQC14}. Unless $\textrm{NP} = \textrm{PSPACE}$, nonblockingness is computationally easier in this case.

  Let $\A$ be an NFA over $\Sigma$ and $P$ be a projection from $\Sigma^*$. Then $P(\A)$ is a DFA such that $L_m(P(\A)) = P(L_m(\A))$ and $L(P(\A)) = P(L(\A))$, called an {\em observer}; cf.~\cite{CL08,Won04} for a construction. In the worst case, $P(\A)$ has exponentially many states compared to $\A$~\cite{wong98,JiraskovaM12}.
  
  \begin{thm}[One-shared-event DFA modular nonblockingness]\label{np-unary}
    Given $n\ge 2$ nonblocking DFAs $(\A_i)_{i=1}^{n}$ with $\A_i$ over $\Sigma_i$ such that $|\cup_{i\neq j} (\Sigma_i\cap\Sigma_j)|=1$. The problem to decide whether $\overline{L_m(\|_{i=1}^{n} \A_i)} = L(\|_{i=1}^{n} \A_i)$ is NP-complete.
  \end{thm}
  \begin{IEEEproof}
    Let $(\A_i)_{i=1}^{n}$ over $(\Sigma_i)_{i=1}^{n}$ be nonblocking DFAs, and let $a$ be the only event such that $\Sigma_i \cap \Sigma_j =\{a\}$, for every $i\neq j$. Let $P$ be a projection from $(\cup_{i=1}^{n} \Sigma_i)^*$ to $\{a\}^*$. Let $m$ be the maximum number of states of all $\A_i$'s. Then $P(\A_i)$ is a DFA with at most $2^m$ states, hence the composition $\A = \|_{i=1}^{n} P(\A_i)$ has at most $2^{mn}$ states, each of the form $(X_1,\ldots,X_n)$, where $X_i$ is a subset of states of $\A_i$. 
    
    Let $\delta$ denote the transition function of $\A$ and $q_0$ its initial state. Notice that $\A$ is a sequence of transitions possibly with a cycle at the end. Then $\A$ is nonblocking iff there exist $k \le 2^{mn}$ and $k < \ell \le 2^{mn+1}$ such that $\delta(q_0,a^k)$ is an accepting state of $\A$ and either $\delta(q_0,a^{k+1})$ is not defined or $\delta(q_0,a^k) = \delta(q_0,a^\ell)$. We now show how to check this property in nondeterministic polynomial time.
    
    The nondeterministic algorithm guesses $k$ and $\ell$ in binary, requiring at most $mn+1$ digits each. To compute the states $\delta(q_0,a^k)$ and $\delta(q_0,a^\ell)$ in polynomial time, the algorithm proceeds as follows.
    
    Let $\A_i'$ denote the NFA obtained from $\A_i$ by replacing each transition $(s,b,t)$ with the transition $(s,P(b),t)$, and by eliminating the $\eps$-transitions afterwards. This can be computed in polynomial time~\cite{HopcroftU79} and is often used as the middle step in the computation of the observer; namely, it preserves the languages. Then $\A_i'$ is over $\{a\}$ and has the same states as $\A_i$. Let $J_i$ denote the set of all initial states of $\A_i'$; it is computed in polynomial time as the set of all states of $\A_i$ reachable under $\Sigma\setminus\{a\}$ from the initial state of $\A_i$ (it is also the initial state of $P(\A_i)$, that is, $q_0=(J_1,\ldots,J_n)$).
    
    The transition relation of $\A_i'$ can be represented as a binary matrix $M_i$, where for states $s,t$ of $\A_i'$, $M_i[s,t]=1$ iff $(s,a,t)$ is a transition in $\A_i'$. For $k\ge 2$, let $M_i^k$ be the multiplication of $M_i$ with itself $k$ times. Then $M_i^k[s,t]$ is the number of paths of length $k$ from $s$ to $t$ in $\A_i'$~\cite{introToAlgs}. 
    Let $\delta_{\A_i'}$ denote the transition function of $\A_i'$. Then $\delta_{\A_i'}(q_i,a^k) = \{ t \mid M_i^k[q_i,t] > 0 \}$ (if it is empty, the transition is undefined). The size of matrix $M_i^k$ is polynomial in the number of states of $\A_i$ and can be computed in time logarithmic in $k$ by fast matrix multiplication: $M_i^2=M_i\times M_i$, $M_i^4 = M_i^2\times M_i^2$, $M_i^8 = M_i^4 \times M_i^4$, \ldots.
    
    To compute $\delta(q_0,a^{k})$, we compute $M_i^k$, for $i=1,\ldots,n$, in polynomial time. Then the state $\delta(q_0,a^{k})=(\delta_{\A_1'}(J_1,a^k), \allowbreak \ldots, \allowbreak \delta_{\A_n'}(J_n,a^k))$ and it is marked iff every $\delta_{\A_i'}(J_i,a^k)$ contains a marked state of $\A_i'$.
    It should now be clear how to check, in polynomial time, that either $\delta(q_0,a^{k+1})$ is not defined or $\delta(q_0,a^k) = \delta(q_0,a^\ell)$, cf. also Example~\ref{ex5} and Remark~\ref{rem2}.
    
    To show NP-hardness, we reduce 3CNF to our problem and use the construction of~\cite{StockmeyerM73}. Let $\varphi$ be a formula in 3CNF (see footnote~\ref{ft1} on page \pageref{ft1}) with $n$ distinct variables and $m$ clauses, and let $C_k$ be the set of literals in the $k$th clause, $1 \le k \le m$. The assignment to the variables is represented as a binary vector of length $n$. Let $p_1,\ldots,p_n$ denote the first $n$ prime numbers. For a natural number $z$ congruent with 0 or 1 modulo $p_i$, for all $i=1,\ldots,n$, $z$ satisfies $\varphi$ if the assignment $(z \bmod p_1,\ldots, z \bmod p_n)$ satisfies $\varphi$. 
    
    For $u=1,\ldots,n$ and $j=2,\ldots,p_u-1$, let $\B'_{u,j}$ denote a nonblocking DFA such that 
    \[
      L_m(\B'_{u,j}) = 0^j\cdot (0^{p_u})^*\,.
    \]
    Then $\cup_{u=1}^{n} \cup_{j=2}^{p_u-1} L_m(\B'_{u,j}) = \{ 0^z \mid \exists u \le n, z \not\equiv 0 \bmod p_u \text{ and } z \not\equiv 1 \bmod p_u \}$ is the set of all natural numbers that {\em do not encode an assignment} to the variables. 
    
    For each clause $C_k$, we construct a nonblocking DFA $\B'_k$ such that if $0^z \in L_m(\B'_k)$ and $z$ is an assignment, then $z$ does not assign value 1 to any literal in $C_k$. For example, if $C_k = \{x_{r}, \neg x_{s}, x_{t}\}$, for $1 \le  r,s,t \le n$ and $r,s,t$ distinct, let $z_k$ be the unique integer such that $0\le z_k < p_rp_sp_t$, $z_k \equiv 0 \bmod p_r$, $z_k \equiv 1 \bmod p_s$, and $z_k \equiv 0 \bmod p_t$. Then
    \[
      L_m(\B'_k) = 0^{z_k} \cdot (0^{p_rp_sp_t})^*\,.
    \]
    
    Let $\B_1,\ldots,\B_\ell$ denote all the DFAs $\B'_{u,j}$ and $\B'_k$ constructed above, and let $\A_i$ denote $\B_i$ with the sets of marked and non-marked states exchanged, that is, $L_m(\A_i) = 0^* \setminus L_m(\B_i)$. Note that all $\B_i$ and $\A_i$ are nonblocking and their generated languages are $0^*$.
    
    Now, $\varphi$ is satisfiable if and only if there exists $z$ such that $z$ encodes an assignment to $\varphi$, i.e., $0^z \notin \cup_{u=1}^{n} \cup_{j=2}^{p_u-1} L_m(\B'_{u,j})$, and $z$ satisfies every clause $C_k$, that is, $0^z \notin L_m(\B'_k)$ for all $k=1,\ldots,m$. This is iff $ 0^z \in \cap_{i=1}^{\ell} L_m(\A_i) = L_m(\|_{i=1}^{\ell} \A_i)$.
    
    We show that $\|_{i=1}^{\ell} \A_i$ is nonblocking iff $L_m(\|_{i=1}^{\ell} \A_i) \neq \emptyset$. 
    
    If $L_m(\|_{i=1}^{\ell} \A_i) = \emptyset$, then $\|_{i=1}^{\ell} \A_i$ is blocking, because $\eps \in L(\|_{i=1}^{\ell} \A_i)$.
    
    If $0^z \in L_m(\|_{i=1}^{\ell} \A_i)$, then $z$ satisfies $\varphi$. For a natural number $c$, the number $z+c\cdot \Pi_{i=1}^{n} p_i$ also satisfies $\varphi$: indeed, if $z \equiv x_i \bmod p_i$, then $(z + c\cdot \Pi_{i=1}^{n} p_i) \equiv x_i \bmod p_i$, for all $i$. Thus, $0^z (0^{\Pi_{i=1}^{n} p_i})^* \subseteq L_m(\|_{i=1}^{\ell} \A_i)$. Since for every $0^s \in L(\|_{i=1}^{\ell} \A_i)$, there exists $c$ such that $s \le z + c\cdot \Pi_{i=1}^{n} p_i$, we have that $\|_{i=1}^{\ell} \A_i$ is nonblocking.
  \end{IEEEproof}

  \begin{remark}
    If the number of DFAs in Theorem~\ref{np-unary} is at most $k$, for a constant $k$, the problem is NL-complete. The membership in NL is by Theorem~\ref{pspace-complete01} and NL-hardness by Theorem~\ref{thm1}. 
  \end{remark}
  
  \begin{example}\label{ex5}
    We illustrate the polynomial computation used in the proof of Theorem~\ref{np-unary} for $n=1$. Its generalization to $n>1$ is straightforward.
    Let $\A_1=(\{1,2,3,4\}, \allowbreak \{a,b\}, \allowbreak \{(1,a,2), \allowbreak (2,a,1),(2,b,3), \allowbreak (3,a,4), \allowbreak (4,a,1)\}, \allowbreak 1, \allowbreak \{1\})$ be a DFA, and let $\A_1'$, depicted in Fig.~\ref{fig7}, 
    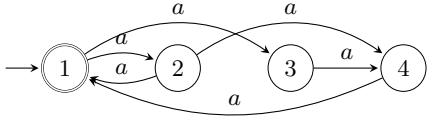
\begin{figure}[h]
      \centering
      \begin{tikzpicture}[baseline,auto,->,>=stealth,shorten >=1pt,node distance=1.5cm,
        state/.style={ellipse,minimum size=6mm,very thin,draw=black,initial text=},
        every node/.style={font=\small}]
        \node[state,initial,accepting]  (1) {$1$};
        \node[state]                    (2) [right of=1] {$2$};
        \node[state]                    (3) [right of=2] {$3$};
        \node[state]                    (4) [right of=3] {$4$};
        \path
          (1) edge[bend left=20] node {$a$} (2)
          (2) edge[bend left=20] node[above] {$a$} (1)
          (1) edge[bend left=35] node {$a$} (3)
          (2) edge[bend left=35] node {$a$} (4)
          (3) edge node {$a$} (4)
          (4) edge[bend left=25] node[above] {$a$} (1)
          ;
      \end{tikzpicture}
      \caption{The NFA $\A_1'$ of Example~\ref{ex5}}
      \label{fig7}
    \end{figure}
    be the NFA obtained from $\A_1$ by renaming $b$-transitions to $\eps$-transitions, and by the elimination of $\eps$-transitions afterwards. Then the $4\times 4$ transition matrix $M_1$ and its $4$th power $M_1^{4}$ are
    \[
      M_1 =
      \begin{pmatrix} 
        0 & 1 & 1 & 0  \\
        1 & 0 & 0 & 1  \\
        0 & 0 & 0 & 1  \\
        1 & 0 & 0 & 0  
      \end{pmatrix}
      \qquad
      M_1^{4} =
      \begin{pmatrix} 
        1 & 2 & 2 & 2  \\
        3 & 1 & 1 & 2  \\
        1 & 0 & 0 & 2  \\
        2 & 1 & 1 & 0  
      \end{pmatrix}\,.
    \]
    The reader may verify that $\delta(\{1\},a^4)=\{1,2,3,4\}$, where $\delta$ is the transition relation of the observer $P(\A_1)$.
  \end{example}
  
  \begin{remark}\label{rem2}
    The number $M_i^k[s,t]$ represents the number of paths from state $s$ to state $t$ of length $k$. This information is not important for us. The information we need is whether there is a path, i.e., $M_i^k[s,t]>0$, or not, i.e., $M_i^k[s,t]=0$. The numbers $M_i^k[s,t]$ may become large and affect thus the complexity. To keep the complexity polynomial (the numbers small), the $+$ operation in the definition of matrix multiplication is replaced by $\max$ operation. This minor trick keeps the matrices $M_i^k$ binary, while providing the same information~\cite{introToAlgs}.
  \end{remark}

\section{Conclusion}\label{practical}
  The theoretical results do not seem very optimistic. However, there are techniques to reduce the size of an automaton, which allows to handle large automata that appear in practical applications. A well-known technique is the BDD diagrams~\cite{Bryant86}. Another technique is the state-tree structures~\cite{MaW06} or the method using extended finite-state machines and abstractions~\cite{MohajeraniMF16}.
  
\section*{Acknowledgment}
  The author gratefully acknowledges very useful suggestions and comments of the anonymous referees.

\bibliographystyle{IEEEtran}
\bibliography{nonblock}

\begin{thebibliography}{10}
\providecommand{\url}[1]{#1}
\csname url@samestyle\endcsname
\providecommand{\newblock}{\relax}
\providecommand{\bibinfo}[2]{#2}
\providecommand{\BIBentrySTDinterwordspacing}{\spaceskip=0pt\relax}
\providecommand{\BIBentryALTinterwordstretchfactor}{4}
\providecommand{\BIBentryALTinterwordspacing}{\spaceskip=\fontdimen2\font plus
\BIBentryALTinterwordstretchfactor\fontdimen3\font minus
  \fontdimen4\font\relax}
\providecommand{\BIBforeignlanguage}[2]{{%
\expandafter\ifx\csname l@#1\endcsname\relax
\typeout{** WARNING: IEEEtran.bst: No hyphenation pattern has been}%
\typeout{** loaded for the language `#1'. Using the pattern for}%
\typeout{** the default language instead.}%
\else
\language=\csname l@#1\endcsname
\fi
#2}}
\providecommand{\BIBdecl}{\relax}
\BIBdecl

\bibitem{MohajeraniMF14}
S.~Mohajerani, R.~Malik, and M.~Fabian, ``A framework for compositional
  synthesis of modular nonblocking supervisors,'' \emph{IEEE Trans. Autom.
  Control}, vol.~59, pp. 150--162, 2014.

\bibitem{Malikwodes2016}
R.~Malik, ``Programming a fast explicit conflict checker,'' in \emph{WODES},
  2016, pp. 438--443.

\bibitem{CL08}
C.~Cassandras and S.~Lafortune, \emph{Introduction to discrete event systems},
  2nd~ed.\hskip 1em plus 0.5em minus 0.4em\relax Springer, 2008.

\bibitem{papadimitriou}
C.~Papadimitriou, \emph{Computational Complexity}.\hskip 1em plus 0.5em minus
  0.4em\relax Addison-Wesley, 1994.

\bibitem{sipser}
M.~Sipser, \emph{Introduction to the theory of computation}, 2nd~ed.\hskip 1em
  plus 0.5em minus 0.4em\relax Thompson Course Technology, 2006.

\bibitem{GareyJ79}
M.~Garey and D.~Johnson, \emph{Computers and Intractability: A Guide to the
  Theory of NP-Completeness}.\hskip 1em plus 0.5em minus 0.4em\relax W. H.
  Freeman, 1979.

\bibitem{Tarjan72}
R.~Tarjan, ``Depth-first search and linear graph algorithms,'' \emph{{SIAM} J.
  Comput.}, vol.~1, pp. 146--160, 1972.

\bibitem{HuntR78}
H.~B. {Hunt III} and D.~J. Rosenkrantz, ``Computational parallels between the
  regular and context-free languages,'' \emph{{SIAM} J. Comput.}, vol.~7, pp.
  99--114, 1978.

\bibitem{RohloffL05}
K.~Rohloff and S.~Lafortune, ``{PSPACE}-completeness of modular supervisory
  control problems,'' \emph{Discrete Event Dyn. Syst.}, vol.~15, pp. 145--167,
  2005.

\bibitem{HolzerK11}
M.~Holzer and M.~Kutrib, ``Descriptional and computational complexity of finite
  automata -- {A} survey,'' \emph{Inform. and Comput.}, vol. 209, pp. 456--470,
  2011.

\bibitem{BrandinW1994}
B.~Brandin and W.~Wonham, ``Supervisory control of timed discrete-event
  systems,'' \emph{IEEE Trans. Autom. Control}, vol.~39, pp. 329--342, 1994.

\bibitem{SchafaschekQC14}
G.~Schafaschek, M.~Queiroz, and J.~Cury, ``Local modular supervisory control of
  timed discrete-event systems,'' in \emph{WODES}, 2014, pp. 271--277.

\bibitem{Won04}
W.~Wonham, ``Supervisory control of discrete-event systems,'' 2009.

\bibitem{wong98}
K.~Wong, ``On the complexity of projections of discrete-event systems,'' in
  \emph{WODES}, 1998, pp. 201--206.

\bibitem{JiraskovaM12}
G.~Jir{\'{a}}skov{\'{a}} and T.~Masopust, ``On a structural property in the
  state complexity of projected regular languages,'' \emph{Theoret. Comput.
  Sci.}, vol. 449, pp. 93--105, 2012.

\bibitem{HopcroftU79}
J.~E. Hopcroft and J.~D. Ullman, \emph{Introduction to Automata Theory,
  Languages and Computation}.\hskip 1em plus 0.5em minus 0.4em\relax
  Addison-Wesley, 1979.

\bibitem{introToAlgs}
T.~Cormen, C.~Leiserson, R.~Rivest, and C.~Stein, \emph{Introduction to
  Algorithms}, 3rd~ed.\hskip 1em plus 0.5em minus 0.4em\relax {MIT} Press,
  2009.

\bibitem{StockmeyerM73}
L.~Stockmeyer and A.~Meyer, ``Word problems requiring exponential time:
  Preliminary report,'' in \emph{{STOC}}, 1973, pp. 1--9.

\bibitem{Bryant86}
R.~Bryant, ``Graph-based algorithms for boolean function manipulation,''
  \emph{{IEEE} Trans. Comput.}, vol.~35, pp. 677--691, 1986.

\bibitem{MaW06}
C.~Ma and W.~Wonham, ``Nonblocking supervisory control of state tree
  structures,'' \emph{{IEEE} Trans. Autom. Control}, vol.~51, pp. 782--793,
  2006.

\bibitem{MohajeraniMF16}
S.~Mohajerani, R.~Malik, and M.~Fabian, ``A framework for compositional
  nonblocking verification of extended finite-state machines,'' \emph{Discrete
  Event Dyn. Syst.}, vol.~26, pp. 33--84, 2016.

\end{thebibliography}

\end{document}